\documentclass[aps,prl,twocolumn,superscriptaddress,amsmath,amssymb,floatfix]{revtex4-2}
\usepackage{graphicx}
\usepackage{dcolumn}
\usepackage{bm}
\usepackage{xcolor}

\newcommand{\ab}{\textcolor{black}}

\newcommand{\abo}{\textcolor{black}}
\newcommand{\jn}{\textcolor{black}}

\newcommand{\rev}{\textcolor{black}} 

\newcommand{\ompi}{\omega_\mathrm{pi}}

\newcommand{\omci}{\Omega_\mathrm{i}}

\newcommand{\ms}{M_\mathrm{s}}

\newcommand{\ma}{M_\mathrm{A}}
\newcommand{\mi}{m_\mathrm{i}}
\newcommand{\me}{m_\mathrm{e}}
\newcommand{\lse}{\lambda_\mathrm{se}}
\newcommand{\lsi}{\lambda_\mathrm{si}}
\newcommand{\vsh}{v_\mathrm{sh}}

\def\ee{\end{equation}}
\def\be{\begin{equation}}

\begin{document}

\preprint{APS/123-QED}

\title{Magnetic field amplification by the Weibel instability at \\ planetary and astrophysical high-Mach-number shocks}

\author{Artem Bohdan}
\email{artem.bohdan@desy.de}
\affiliation{DESY, DE-15738 Zeuthen, Germany}

\author{Martin Pohl}
\affiliation{DESY, DE-15738 Zeuthen, Germany}
\affiliation{Institute of Physics and Astronomy, University of Potsdam, DE-14476 Potsdam, Germany}

\author{Jacek Niemiec}
\affiliation{Institute of Nuclear Physics Polish Academy of Sciences, PL-31342 Krakow, Poland}

\author{Paul J. Morris}
\affiliation{DESY, DE-15738 Zeuthen, Germany}

\author{Yosuke Matsumoto}
\affiliation{Department of Physics, Chiba University, 1-33 Yayoi-cho, Inage-ku, Chiba 263-8522, Japan}

\author{Takanobu Amano}
\affiliation{Department of Earth and Planetary Science, the University of Tokyo, 7-3-1 Hongo, Bunkyo-ku, Tokyo 113-0033, Japan}

\author{Masahiro Hoshino}
\affiliation{Department of Earth and Planetary Science, the University of Tokyo, 7-3-1 Hongo, Bunkyo-ku, Tokyo 113-0033, Japan}

\author{Ali Sulaiman}
\affiliation{Department of Physics and Astronomy, University of Iowa, IA, USA}

\date{\today}

\begin{abstract}
Collisionless shocks are ubiquitous in the Universe and often associated with strong magnetic field.  Here we use large-scale particle-in-cell simulations of non-relativistic perpendicular shocks in the high-Mach-number regime to study the amplification of magnetic field within shocks. 
\ab{The magnetic field is amplified at the shock transition due to the ion-ion two-stream Weibel instability. The normalized magnetic-field strength strongly correlates with the Alfv\'enic Mach number. Mock spacecraft measurements derived from PIC simulations are fully consistent with those taken in-situ at Saturn's bow shock by the Cassini spacecraft.}
\end{abstract}

\maketitle




Collisionless shocks are ubiquitous in the Universe, and they are observed in planetary systems, supernova remnants (SNRs), jets of active galactic nuclei, galaxy clusters, etc. In contrast to fluid shock waves, where dissipation at the shock front is mediated by binary collisions, collisionless shocks are shaped by collective particle interactions with interaction length much shorter than the collisional mean free path \citep{1966RvPP....4...23S,1971swcp.book.....T}.
Collisionless shocks are usually magnetized, and magnetic fields play a key role in their physics. The jump condition for the magnetic field \citep{2020ApJ...900..111B} and the internal shock structure \citep{2009A&ARv..17..409T} strongly depends on the shock obliquity.  Magnetic field turbulence near the shock is a key ingredient of diffusive shock acceleration (DSA, \cite{1977ICRC...11..132A,1977DoSSR.234.1306K,1983RPPh...46..973D,1978MNRAS.182..147B,1978MNRAS.182..443B,1978ApJ...221L..29B}) and also shapes non-thermal X-ray emission.
Amplified magnetic fields (at scales much larger \jn{than} the upstream ion gyroradius) have been inferred from observations of SNRs through the detection of non-thermal X-ray rims \cite{2003ApJ...584..758V,2004AdSpR..33..376B,2005ApJ...626L.101P,2012A&ARv..20...49V}, fast temporal variability of X-ray hot spots \cite{2007Natur.449..576U}, and the $\gamma$-ray/X-ray flux ratio \cite{2011ApJ...730L..20A}. We know various possible mechanisms for magnetic field amplification at these scales: cosmic-ray driven nonresonant modes \cite{2004MNRAS.353..550B,2005MNRAS.358..181B}, fluid vorticity downstream of the shock seeded by upstream density inhomogeneities \cite{2007ApJ...663L..41G,2013ApJ...770...84F}, cosmic ray pressure-driven magnetic field amplification \cite{2012MNRAS.425.2277D,2014MNRAS.444..365D}, and also inverse cascading of relatively short-scale Alfv\'en waves \cite{2007ApJ...654..252D}. 

Here we study magnetic field amplification on scales smaller than the upstream ion gyroradius at high-Mach-number quasi-perpendicular shocks. 
In-situ measurements by the Cassini spacecraft \citep{2015PhRvL.115l5001S,2016JGRA..121.4425S} reveal the detailed magnetic field structure of Saturn's bow shock with resolution below the ion gyroradius. The Alfv\'enic Mach number of this shock can reach values around 200 which is similar to that of SNR shocks.  \citep{2016JGRA..121.4425S} demonstrated that the normalized overshoot magnetic-field strength displays a strong positive correlation with $\ma$ across the entire range of measured $\ma$. 
Particularly strong amplification is observed at shocks at which shock self-reformation is evident \citep{2015PhRvL.115l5001S}. Reasons for such behaviour are unknown and \jn{they are} the objective of our study.


Leroy's calculations \cite{1983PhFl...26.2742L} for perpendicular shocks combined with hybrid simulations suggest that the overshoot magnetic-field strength ($B_{\rm over}$) can be estimated as 
\be
B_{\rm over} \approx 0.4 B_0 \ma^{7/6},
\label{B_over} 
\ee
where $B_0$ is the upstream field strength.
The prefactor $0.4$ was determined with simulations \cite{1986JGR....91.8805Q,Matsumoto2012}. In this model the magnetic-field amplification is associated only with plasma compression, and multidimensional effects may not be accounted for. 
However 3D \cite{Matsumoto2017} and some 2D \cite{2010ApJ...721..828K,Matsumoto2015} PIC simulations of quasi-perpendicular high-$\ma$ shocks demonstrate strong amplification of the upstream magnetic field due to the ion-ion filamentation/Weibel instability \citep{1959PhRvL...2...83W,1959PhFl....2..337F}, which results from the interaction of upstream and shock-reflected ions. 
The mediation of high-$\ma$ shocks by the Weibel instability is also confirmed by laboratory experiments \cite{2020NatPh..16..916F} and in-situ measurements of the Earth's bow shock at $\ma \simeq 39$ \cite{2017ApJ...836L...4S}. 
In this letter we discuss a mechanism of magnetic-field amplification that is based on a realistic description of perpendicular nonrelativistic high-$\ma$ shocks and can explain the correlation between field strength and $\ma$ observed with Cassini at Saturn's bow shock.

\begin{table}
      \caption{Parameters of simulation runs. Listed are: the ion-to-electron mass ratio, $\mi/\me$, the Alfv\'enic and sonic Mach number, $\ma$ and $\ms$, the electron plasma beta, $\beta_{\rm e}$. Some values are shown separately for the \emph{left} (runs *1) and the \emph{right} (runs *2) shock. Results for runs marked by a '$^\dagger$' are not discussed in this letter because of the strong numerical noise at the shock upstream. All runs use the \emph{in-plane} magnetic field configuration, $\varphi=0^o$.
         \label{table-param}}
         \begin{ruledtabular}
\begin{tabular}{ccccccc}
Runs  & $\mi/\me$ & $\ma$ & \multicolumn{2}{c}{$\ms$} & \multicolumn{2}{c}{$\beta_{\rm e}$} \\
 & & & $^*1$ & $^*2$ & $^*1$ & $^*2$ \\
\noalign{\smallskip}
\hline
\noalign{\smallskip}
A1, A2   & 50  &   22.6  & 1104 & 35  & $5 \cdot 10^{-4}$ & 0.5 \\
B1, B2   & 100 &   31.8  & 1550 & 49  & $5 \cdot 10^{-4}$ & 0.5 \\
C1, C2   & 100 &   46    & 2242 & 71  & $5 \cdot 10^{-4}$ & 0.5 \\
D1, D2   & 200 &   32    & 1550 & 49  & $5 \cdot 10^{-4}$ & 0.5 \\
E1, E2   & 200 &   44.9  & 2191 & 69  & $5 \cdot 10^{-4}$ & 0.5 \\
F1, F2   & 400 &   68.7  & 3353 & 106 & $5 \cdot 10^{-4}$ & 0.5 \\
\ab{G1, G2}   & 50 &   68.7  & 3353 & 106 & $5 \cdot 10^{-4}$ & 0.5 \\
H1$^\dagger$, H2   & 50  &   100   & 4870 & 154 & $5 \cdot 10^{-4}$ & 0.5 \\
I1$^\dagger$, I2   & 50  &   150   & 7336 & 232 & $5 \cdot 10^{-4}$ & 0.5 
\end{tabular}
\end{ruledtabular}
\end{table}


To tackle this issue  we use 2D PIC simulations with an \emph{in-plane} magnetic-field configuration which permits a good approximation of realistic 3D shocks \cite{Bohdan2017,Matsumoto2017}.
We perform shock simulations using an optimized fully-relativistic electromagnetic 2D code with MPI parallelization developed from TRISTAN \citep{Buneman1993,2008ApJ...684.1174N,10.1007/978-3-319-78024-5_15}. 
Shocks are initialized with a modified flow-flow method \cite{2016ApJ...820...62W}. The collision of two counterstreaming electron-ion plasma flows, each described with 20 particles per cell per species, spawns two independent shocks propagating in opposite directions. The inflow speed of two beams is $v_{\rm L}=v_{\rm R}=v_{\rm 0}=0.2c$. 
The plasma temperature for two flows differs by a factor of $1000$, therefore \emph{electron} plasma beta (the ratio of the electron plasma pressure to the magnetic pressure) is $5 \cdot 10^{-4}$ and 0.5 for the \emph{left} (runs *1) and the \emph{right} (runs *2) shocks, respectively. 

\abo{
The large-scale magnetic field, $\bf B_0$, is perpendicular to the shock normal ($\theta_{Bn}=90^o$) and lies in the simulation plane (the \emph{in-plane} configuration, $\varphi=0^o$). The adiabatic index is $\Gamma_{\rm ad}=5/3$, the shock compression ratio is about 4, and the shock speed in the \emph{upstream} frame is $\vsh=0.263c$. 
The Alfv\'en velocity is $v_{\rm A}=B_{\rm 0}/\sqrt{\mu_{\rm 0}(N_{\rm e}\me+N_{\rm i}\mi)}$, where $\mu_{\rm 0}$ is the vacuum permeability; $N_{\rm i}$ and $N_{\rm e}$ are the ion and electron number density. The sound speed reads $c_{\rm s}=(\Gamma_{\rm ad} k_{\rm B}T_{\rm i}/\mi)^{1/2}$, where $k_{\rm B}$ is the Boltzmann constant and $T_{\rm i}$ is the ion temperature.
The Alfv\'enic, $\ma=\vsh/v_{\rm A}$, and sonic, $\ms=\vsh/c_{\rm s}$, Mach numbers of the shocks are defined in the conventional \emph{upstream} frame \ab{(Table 1)}.}

\abo{
The ratio of the electron plasma frequency, $\omega_{\rm pe}=\sqrt{e^2N_{\rm e}/\epsilon_{\rm 0}\me}$, to the electron gyrofrequency, $\Omega_{\rm e}=eB_0/\me$, \jn{is} in the range $\omega_{\rm pe}/\Omega_{\rm e}=8.5-80$. 
Here, $e$ is the electron charge, and $\epsilon_0$ is the vacuum permittivity. The temporal and spatial resolutions are $\delta t=\frac{1}{40}\omega_{\rm pe}^{-1}$ and $\Delta = \frac{1}{20}\lse$, where $\lse$ is the electron skin depth. \ab{The transverse box size is $L_{\rm y}=(8-24)\lsi$, where $\lsi=\sqrt{\mi/\me}\lse$ is the ion skin depth}. The simulation time is about $T \approx 8 \Omega_{\rm i}^{-1}$, where $\Omega_{\rm i}=eB_0/\mi$. }

Our simulations cover a wide range of physical parameters: $\ma = 22.6 - 150$, $\mi/\me=50-400$ and $\beta_{\rm e,R}=5 \cdot 10^{-4}-0.5$. Hence, we can compare our simulation results with data for Saturn's bow shock for $\ma \geqslant 20$. 

\begin{figure}[!t]
\centering
\includegraphics[width=0.99\linewidth]{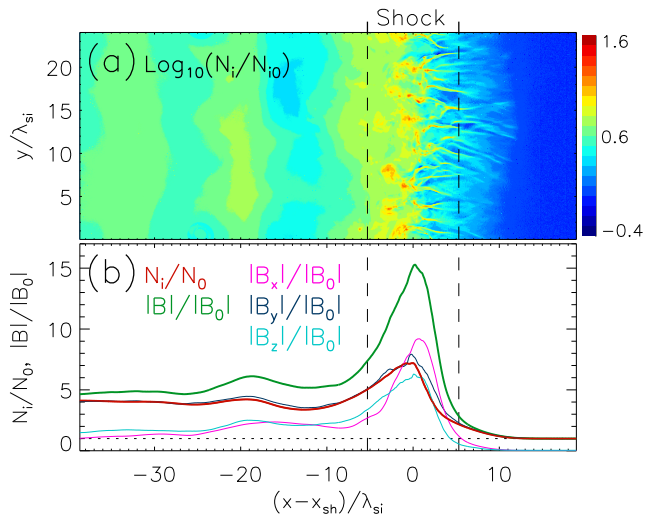}
\caption{Density  and magnetic field in run B2. Panel (a): ion density in logarithmic units. Panel (b): red line - the profile of normalized ion density, green line - the profile of normalized magnetic field, \ab{magenta line - $B_{\rm x}/B_0$, dark blue line - $B_{\rm y}/B_0$, light blue line - $B_{\rm z}/B_0$}. Profiles are calculated in the shock reference frame and averaged over the shock reformation cycle. The shock region is marked by dashed lines. $x_{\rm sh}$ is the shock position.}
\label{dens_B2_prof}
\end{figure}

Figure~\ref{dens_B2_prof}(a) shows an electron-density map of the fully developed shock from run B2. The shock position, $x_{\rm sh}$, is defined as position of the shock overshoot. Buneman waves are visible as small-scale density ripples at $x-x_{\rm sh} \approx (8-12)\lsi$. The Weibel instability is represented by density filaments at $x-x_{\rm sh}\approx (2-10) \lsi$. The downstream region is at $x-x_{\rm sh} < -5  \lsi$. This structure is representative for all runs \ab{and for the high-$\ma$ regime in general \cite{2010ApJ...721..828K,Matsumoto2015,Bohdan2017,Bohdan2019a,Bohdan2019b,Bohdan2020a,Bohdan2020b}.} \rev{Earlier linear analysis  \cite{2010ApJ...721..828K} and its adaptation to our study \cite{Bohdan2020a} both indicate that high-$\ma$ shocks are Weibel-instability mediated.}

Figure~\ref{dens_B2_prof}(b) displays the density and magnetic-field profiles at the shock transition of run B2, averaged in time over two cycles of shock reformation. The plasma compression reaches $N_{\rm over}/N_0 \approx 7$ at the shock overshoot in all simulations, which is not in line with Leroy's model, \ab{where $N_{\rm over}$ depends on $\ma$}. The field strength increases twice as much, indicating substantial noncompressional magnetic-field amplification. 

\ab{
The $B_{\rm y}$ profile almost coincides with that expected for simple compression of $B_{\rm y}$ according to the density profile. 
The modest increase of $B_{\rm y}$ around $(x-x_{\rm sh})/\lsi \approx (0-3)$ is due to magnetic reconnection which turns $B_{\rm x}$ into $B_{\rm y}$ \rev{when magnetic loops elongated in $x$-direction break up into chains of magnetic vortices \cite{Bohdan2020a}}. As expected, $B_{\rm x}$ and $B_{\rm z}$ grow due to folding of magnetic field by the Weibel modes whose wave vector is perpendicular to the relative velocity of shock-reflected and incoming upstream ions \cite{2010ApJ...721..828K,Matsumoto2015}.} 
Further straightening of magnetic-field lines leads to convergence of the density and magnetic field profiles at the shock downstream.





We define the shock region as a sector of width $L_{\rm sh}=r_{\rm gi,up}/3$ centered at $x_{\rm sh}$, where $r_{\rm gi,up}=\ma \lsi$. The numerical coefficient is chosen to match the shock width and the average ion gyroradius at the shock transition layer; its exact value has little, if any, impact on the results discussed here. The shock region for run B2 is marked with dashed lines in Figure~\ref{dens_B2_prof}. 

\begin{figure}[!t]
\centering
\includegraphics[width=0.99\linewidth]{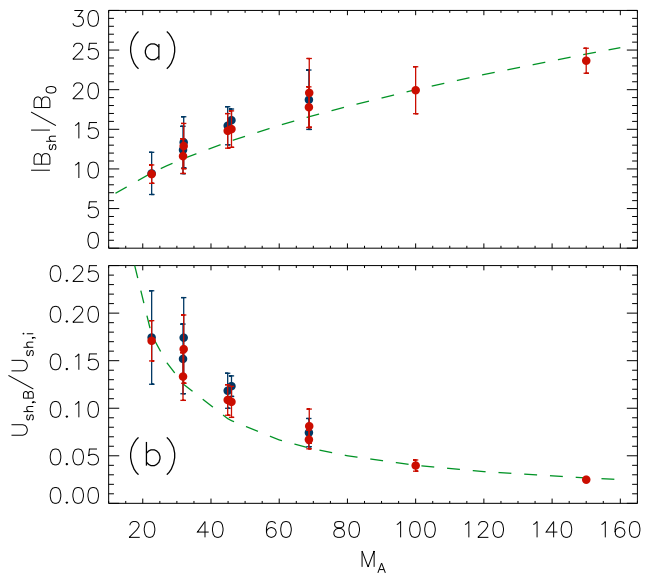}
\caption{The normalized magnetic-field strength (panel (a)) and the magnetic energy density normalized by the upstream ion energy density (panel (b)), both evaluated in the shock region defined in Fig.~\ref{dens_B2_prof}. The blue and red color corresponds to \emph{left} ($\beta=5 \cdot 10^{-4}$) and \emph{right} ($\beta=0.5$) shocks, respectively. 
The green dotted line in panel (a) reflects $|B_{\rm sh}|/B_0 = 2 \sqrt{\ma}$, and that in panel (b) shows $U_{\rm sh,B}/U_{\rm sh,i} = 4 \ma^{-1}$.}
\label{energy_Bfield}
\end{figure}

In Figure~\ref{energy_Bfield} we present the amplitude (panel a) and energy density (panel b) of the magnetic field in the shock region, averaged over the shock self-reformation cycle and with error bars reflecting the level of temporal variation.
The normalized field strength, $|B_{\rm sh}|/B_0$, grows with increasing $\ma$. The Weibel growth rate is about $\Gamma \approx 0.1 \ompi$ regardless of the shock parameters \cite{Bohdan2020a}. Shock-self reformation limits the time available for the Weibel instability to develop to about $\omci^{-1}$, implying that the number \jn{of} exponential growth cycles is proportional to $\ma$ for a given shock speed. Exponential growth of the amplitude of Weibel filaments is not observed though even at low $\ma$. In fact, the Weibel instability quickly becomes nonlinear, and the magnetic-field strength defies an analytical derivation. Here we can only estimate it as (green line in Fig~\ref{energy_Bfield}(a))
\be
|B_{\rm sh}| \approx 2 \sqrt{\ma}\, B_0 \ .
\label{bshock} 
\ee
The normalized energy density of the magnetic field can be expressed as
\be
\frac{U_{\rm sh,B}}{U_{\rm sh,i}} = \frac{B^2}{\mu_0 N_{\rm i}\mi\vsh^2}\approx \frac{4}{\ma},
\label{bshockene} 
\ee
which is a descending trend (green line in Fig~\ref{energy_Bfield}(b)), that well reproduces the energy density observed in the simulations. 
A~lower limit for the normalized magnetic energy density should be provided at very high $\ma$ or unmagnetized shocks. For the latter the fraction of magnetic energy in the shock region is about $U_{\rm B} = 0.006\, U_{\rm sh,i}$ \cite{2008ApJ...681L..93K}, which with Eq.~\ref{bshockene} is expected at $\ma \approx 670$, where  $|B_{\rm sh}| \approx 50\,B_0$.

The magnetic field remains amplified for only a few ion gyroradii behind the shock, and far downstream the field strength is $4B_0$. Our simulation time is too short to fully capture the entire relaxation especially for high $\ma$. The data we have suggest that the length scale of relaxation is roughly proportional to  $|B_{\rm sh}|/|B_{0}|$.

\begin{figure}[!t]
\centering
\includegraphics[width=0.99\linewidth]{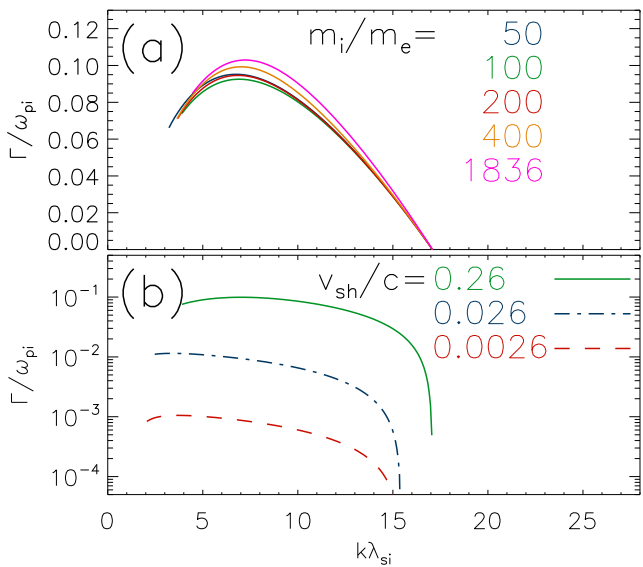}
\caption{Growth rate of Weibel modes for \ab{five mass ratios (a) and} three shock speeds \ab{(b)}.}
\label{Weibel_groth_rate}
\end{figure}

\ab{
We use the analytical description presented in \cite{Bohdan2020a} to clarify the relation between the Weibel growth rate and the choice of plasma parameters, namely, the upstream plasma beta, the mass ratio, and the shock speed. 
Runs *1 and *2 differ by the upstream plasma temperature. At the shock foot, however, the temperature of the plasma constituents is similar on account of partial thermalization, which leads to similar Weibel growth rates. 
Runs that differ only in the mass ratio also show the same magnetic-field amplification level. We use the plasma parameters observed in the shock foot of run F2 to calculate the Weibel instability growth rate for different mass ratios, keeping all kinetic and thermal parameters constant. We find that the growth rate of the most unstable mode remains the same within $\sim 10\%$ margin (Fig.~\ref{Weibel_groth_rate}(a)). 
Therefore we conclude that the upstream plasma beta and the mass ratio do not play a significant role in magnetic-field amplification.}

\ab{We also explore how the behavior of the Weibel instability depends} on the shock speed, which in the simulations is two orders of magnitude higher than at Saturn's bow shock. Figure~\ref{Weibel_groth_rate}(b) shows the Weibel instability growth rates for three values of the shock velocity: $0.26c$, $0.026c$ and $0.0026c$. The last case with $\vsh=780~$km/s is very close to the speed of Saturn's bow shock, \ab{which is about 400 km/s \cite{2006JGRA..111.3201A}}. For $\vsh=0.26c$, we use plasma parameters from run F2. For the two other cases we \ab{accordingly rescale the velocity and the temperature of the plasma flow.} To be noted from Figure~\ref{Weibel_groth_rate}(b) is that the normalized peak growth rate is proportional to the shock speed:
\be
\Gamma_{\rm max} \propto \vsh\,{\ompi}\quad\mathrm{or}\quad 
\Gamma_{\rm max} \propto \ma\,{\omci}
\label{gamma2} 
\ee
This finding matches the result of earlier, simplified calculations \cite{2009A&ARv..17..409T}. 
Eq.~\ref{gamma2} shows that the number of exponential growth cycles available for Weibel modes scales inversely with the Mach number, whatever the shock speed. Therefore $\ma$ is the only \emph{upstream} parameter that defines magnetic-field amplification at the shock transition.

The intrinsic shock dynamics also affects the magnetic-field amplification level. \cite{2015PhRvL.115l5001S} showed that 16 shocks out of 54 shock crossings undergo shock reformation, and the \ab{measured $B_{\rm max}/B_0$ ($B_{\rm max}$ is the maximal magnetic field measured during a shock crossing by the spacecraft)} at these shocks is 1.42 times that at the other 38 shocks. This behaviour is likely explained by the differences in ion reflection at the shock ramp between reforming and non-reforming shocks. With shock self-reformation, the ion reflection rate is time dependent and swings periodically \cite{2016ApJ...820...62W}, reaching larger values than for a non-reforming shock where the ion reflection rate is steady. This results in a stronger magnetic-field amplification in reforming shocks, on account of the higher growth rate of Weibel modes and stronger plasma compression at the shock ramp.
Therefore, $B_{\rm max}/B_0$ is higher for shocks at which shock reformation is observed. In all of our simulations shock reformation is clearly visible. To properly compare with the full set of in-situ measurements, \ab{which includes both reforming and non-reforming shocks}, we therefore reduce the peak field strength measured in the simulations by a factor of \ab{$1.42/(1.42 n_r + (1-n_r))=1.26$, where $n_r=16/54$ is the fraction of reforming shocks} in the in-situ data of \cite{2015PhRvL.115l5001S}.

\begin{figure}[!t]
\centering
\includegraphics[width=0.99\linewidth]{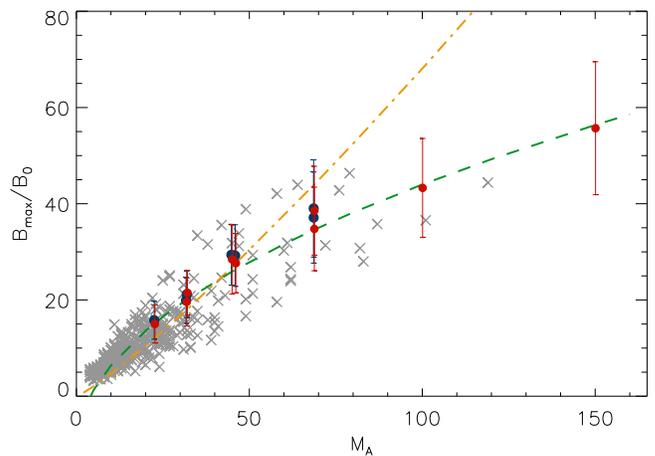}
\caption{Cassini measurements \citep{2016JGRA..121.4425S} indicated by gray crosses and PIC simulation data displayed with blue and red dots for \emph{left} ($\beta=5 \cdot 10^{-4}$) and \emph{right} ($\beta=0.5$) shocks, respectively. The yellow dash-dotted line is an earlier prediction, $B_{\rm over}/B_0 \approx 0.4 \ma^{7/6}\ab{/1.26}$ (cf. Eq.~\ref{B_over})\ab{, corrected for shock reformation}. The green dashed line is the behavior found in our PIC simulations, $B_{\rm max}/B_0 = \ab{5.5} \left( \sqrt{\ma} -2\right)$.}
\label{Bfield}
\end{figure}

The largest set of magnetic-field measurements at Saturn's bow shock \cite{2016JGRA..121.4425S} contains 422 shock crossings during which the shock was quasi-perpendicular, $\theta_{B_n} \geqslant 45^o$, and for which $B_{\rm max}/B_0$ is indicated by gray crosses in Figure~\ref{Bfield}. 
\ab{We derive $B_{\rm max}/B_0$ from PIC simulation data assuming that a virtual spacecraft crosses a simulated shock with a straight trajectory. On the spacecrafts trajectory we calculate $B_{\rm max}/B_0$  and then we average it over all possible shock crossing points} and the speed and flight direction of the virtual spacecraft.
\ab{Hereby we account for both the temporal and the spatial variations of $B_{\rm max}/B_0$.} The results are shown in Figure~\ref{Bfield} as blue and red dots with error bars. Note, that we already applied \ab{to both our results and Leroy's model} the downward correction by the factor $1.26$ that we discussed in the preceding paragraph as compensation for shock reformation.

Figure~\ref{Bfield} demonstrates a good match between in-situ measurement and simulation data. A good fit of the simulation data is shown as green dashed line in Fig.~\ref{Bfield},
\be
\frac{B_{\rm max}}{B_0} = \ab{5.5} \left( \sqrt{\ma} -2\right) ,
\label{eq-bfield} 
\ee
which also well describes the in-situ measurements for $\ma \gtrsim 10$. This is not proof that magnetic fields are defined by Weibel instability at $10 < \ma < 20$, but at least Eq.~\ref{eq-bfield} can be used to estimate the field strength.  
For comparison, the yellow dash-dotted line in Fig.~\ref{Bfield} shows the scaling \jn{of} Eq.~\ref{B_over}, which also was confirmed with recent 2D simulations \cite{Matsumoto2012}. However, 2D simulations cannot always capture realistic shock physics, the \emph{out-of-plane} magnetic field configuration utilized in \cite{Matsumoto2012} misses the Weibel instability, which changes the magnetic field amplification physics compared to our \emph{in-plane} 2D simulations and the 3D simulations of \cite{Matsumoto2017}.
Although Eq.~\ref{B_over} matches the data reasonably well for $\ma < \ab{60}$, even that may be a coincidence because this model relies on simplified 1D shock physics.
In our view, Eq.~\ref{eq-bfield} is a better and physically motivated approximation for $B_{\rm max}/B_0$ at shocks with $\ma \gtrsim 10$.

In addition to the good fit of \ab{$B_{\rm max}/B_0$, the shock reformation period, $T_\mathrm{reform} \approx 1.5 \omci^{-1}$, is the same in our simulations and in the Cassini data \cite{2015PhRvL.115l5001S}. Also the magnetic-field relaxation distance is similar with about one shock width, further suggesting similar physical processes at play in PIC simulations and real bow shocks.}



We find no evidence for magnetic-field amplification by ion beam cyclotron instabilities. They would require more time to develop, $T \gg \omci^{-1}$, and usually these instabilities are observed at quasi-parallel shocks where the shock-reflected ions can move far upstream.

We have established a strong connection between the Weibel instability and magnetic-field amplification at high-$\ma$ shocks. \rev{The results of our PIC simulations are fully consistent with in-situ measurements of Saturn's bow shock.} As $\ma$ is the only relevant parameter, our findings on field amplification inside the shock transition layer should also apply to SNR shocks. Weibel modes can increase the local synchrotron emissivity by a factor $(B_{\rm sh}/B_0)^2$, which may reach a thousand. Larger enhancements arise in the X-ray band beyond the synchrotron peak frequency, but overall the effect is likely unobservable with current facilities \ab{due to low resolution}. However, the interaction of Weibel modes with other amplification processes may introduce significant changes in the shock structure and it should be taken into account in further studies.

Electron pre-acceleration \cite{Bohdan2017,Bohdan2020a} and heating \cite{Bohdan2020b} strongly depend on the structure and strength of the magnetic field. At quasi-perpendicular shocks, where stochastic shock drift acceleration (SSDA) is expected to operate \cite{Matsumoto2017,2019ApJ...874..119K}, strong magnetic field generated by the Weibel instability limits the mean free path and increases the cyclotron frequency of electrons, and so the cut-off energy of SSDA may depend on $\ma$.

Also due to the strong magnetic field at the shock transition \ab{particles} require larger momenta for injection into \rev{classical} DSA, \rev{they repeatedly cross the shock without significant deflection in the shock internal structure. The Larmor radius in the amplified field (Eq.~\ref{bshock}) should then be much larger than the shock width, $r_{\rm sh} \propto r_{\rm gi,up} \propto B_{\rm 0}^{-1}$, which implies for the injection momentum
\be
p_{\rm inj} \propto r_{\rm sh} B_{\rm sh} \propto \sqrt{\ma} ,
\label{ele-inj} 
\ee
at Weibel-mediated shocks.}



\smallskip


\begin{acknowledgements}

The work of J.N. has been supported by Narodowe Centrum Nauki through research project 2019/33/B/ST9/02569. M.P. acknowledges support by DFG through grant PO 1508/10-1. 
\jn{This research was supported by PLGrid Infrastructure. Numerical experiments were conducted on the Prometheus} system at ACC Cyfronet AGH and \jn{the} North-German Supercomputing Alliance (HLRN) under projects bbp00014 and bbp00033.
\end{acknowledgements}

\bibliography{apssamp}

\end{document}